\documentclass[twocolumn,showpacs,preprintnumbers,amsmath,amssymb]{revtex4}
\usepackage{color}
\usepackage{graphicx}
\usepackage{dcolumn}
\usepackage{bm}

\newcommand{\be}[1]{\begin{equation}\label{eq:#1}}
\newcommand{\ee}{\end{equation}}
\newcommand{\bea}{\begin{eqnarray}}
\newcommand{\eea}{\end{eqnarray}}
\newcommand{\bt}{\textbf}

\newcommand{\phd}{\phantom{\dag}}

\newcommand{\up}{^{\phd}}
\newcommand{\noi}{\noindent}
\newcommand{\no}{\nonumber}

\begin{document}
\def\v#1{{\bf #1}}


\title{Nematicity from mixed $\bm{S_{\pm}+d_{x^2-y^2}}$ states in iron-based superconductors}

\author{G. Livanas$^1$}
\author{A. Aperis$^1$}
\author{P. Kotetes$^2$}
\author{G. Varelogiannis$^1$}
\affiliation{$^1$Department of Physics, National Technical University
of Athens, GR-15780 Athens, Greece\\
$^2$Institut f\"{u}r Theoretische Festk\"{o}rperphysik, Karlsruhe Institute of Technology, D-76128 Karlsruhe, Germany}

\vskip 1cm
\begin{abstract}
We demonstrate that in iron-based superconductors (SC), the extended $S_\pm$ SC state {\it coexists} with the $d_{x^2-y^2}$ state under generic conditions. The mixed $S_\pm+d_{x^2-y^2}$
SC is a natural nematic state in which {\it the tetragonal symmetry $C_4$ is broken to $C_2$} explaining puzzling findings of nematic SC in FeSe films \cite{ScienceFeSe}. Moreover, we
report the possibility of a first order transition at low-T from the {\it nematic} $S_\pm+d_{x^2-y^2}$ state to the pure $d_{x^2-y^2}$ state induced by the {\it Zeeman magnetic field}
proposing an original experimental strategy for identifying our mixed nematic state in FeSe films. Extrapolating our findings, we argue that nematicity in non superconducting states
of underdoped and undoped pnictides may reflect mixed $S_\pm+d_{x^2-y^2}$ Density Wave states.

\end{abstract}

\maketitle

Iron-based superconductors (SC) support high cri\-ti\-cal temperatures and unconventional superconducting states like cuprates. However, while in high-T$_c$ hole doped cuprates the
gap is certainly \emph{d-wave}, in iron-based materials the situation is much more complex and still under investigation. In a series of materials like Ba$_{1-x}$K$_x$Fe$_2$As$_2$,
BaFe$_{2-x}$Co$_x$As$_2$, FeTe$_{1-x}$Se$_x$ and K$_x$Fe$_{2-x}$Se$_2$ there is experimental evidence for nodeless superconductivity \cite{nodeless}. Apparently, this is not the
usual isotropic s-wave gap but it is instead the extended s-wave $S_{ext}\equiv S_\pm$ gap changing sign between the electron and hole Fermi surfaces \cite{Spm}. On the other hand, in materials like KFe$_2$As$_2$,
LiFeP, LaOFeP, BaFe(As$_{1-x}$P$_x$)$_2$ and BaFe$_{2-x}$Ru$_x$As$_2$ experiments as diverse as NMR, STM, thermal conductivity and penetration depth measurements all point to the
presence of gap nodes on the FS \cite{nodal,Daghero}. On the theoretical side, unconventional either nodeless or nodal gap structures are usually regarded as evidence of non-phononic
mechanisms \cite{Spm,Daghero,Bernevig}. Nonetheless, we have reported recently \cite{AperisSmallQprb2011} that electron-phonon interaction dominated by small-q processes produces nodeless
$S_\pm$ as well as nodal SC states depending on the doping, including a {\it phonon-driven triplet p-wave state} \cite{AperisSmallQprb2011} possibly observed recently in LiFeAs
\cite{expsTripletLiFeAs}.

Among the iron-based SC, FeSe is the simplest compound representing a prototype for this class of materials on which ideas may be tested. Recent scanning tunneling microscopy and
spectroscopy measurements on high quality FeSe films by Song {\it et al.} \cite{ScienceFeSe} reveal an astonishing feature. Apparently, the tetragonal symmetry C$_4$ is broken to
C$_2$ pointing to a \emph{nematic} SC state, that can certainly not be attributed to the tiny orthorhombic distortion of the FeSe lattice. Nematicity in iron-pnictides is already a
highly debated issue following reports for a possible electronic nematic phase transition \cite{Fradkin} that often coincides and possibly drives \cite{Analytis,Meingast,Boehmer} the
orthorhombic distortion which accompanies the antiferromagnetic transition in undoped and underdoped iron pnictides. Because of the coincidence of nematicity and antiferromagnetism,
it has been suggested that the magnetism itself drives an electronic nematic phase transition \cite{Kivelson,Sachdev,Si,Schmalian}. However, in the case of FeSe there are no such
antiferromagnetic phases involved. As a matter of fact, there have been also proposals for antiferromagnetism-independent-nematicity such as ferro-orbital nematic ordering
\cite{Phillips}.

In the present Letter we introduce a novel approach to the phenomenon of nematicity in pnictides. We demonstrate that S$_\pm$ SC {\it coexists} under generic conditions
with d$_{x^2-y^2}$ SC and the mixed S$_{\pm}$+d$_{x^2-y^2}$ SC state is a \emph{prominent nematic state} capable of explaining the reports of nematic SC in FeSe films
\cite{ScienceFeSe}. Moreover, we point out that at low temperatures, a \emph{Zeeman} field can induce a \emph{first order transition} from the nematic S$_{\pm}$+d$_{x^2-y^2}$ state
to the pure d$_{x^2-y^2}$ state. We therefore propose an experimental strategy for identifying the mixed S$_{\pm}$+d$_{x^2-y^2}$ character of the nematic SC state by repeating the
experiments of Song {\it et al.} \cite{ScienceFeSe} in the presence of {\it in-plane} magnetic fields in which case the Zeeman effect will dominate if the FeSe films are sufficiently
thin \cite{FuldeFilms}. Quite remarkably, we were able to produce self-consistently a Zeeman Field - Temperature phase diagram associated with the mixed nematic phase that exhibits
three distinct SC regions and a \emph{tetracritical} point reminding the one observed in UPt$_3$ \cite{SigristUeda}. Finally, we show that \emph{singlet} mixed SC states that violate
\emph{time-reversal symmetry} (${\cal T}$) are also accessible, exhibiting Zeeman field-induced transitions within the SC phase.

For our generic discussion, we can model qualitatively iron-based SC like FeSe with a {\it minimal and sufficient two-band model} exhibiting a hole pocket around the
$\Gamma$(0,0)-point and an electron pocket around M($\pi$, $\pi$)-point: $\varepsilon_{e}(\bm{k})=t_{1}(\cos k_{x}+\cos k_{y})-t_{2}\cos k_{x}\cos k_{y}+C-\mu$ and
$\varepsilon_{h}(\bm{k})=t_{1}(\cos k_{x}+\cos k_{y})+t_{2}\cos k_{x}\cos k_{y}-C-\mu$ where $\mu$ is the chemical potential and we set $t_1=1$, $t_2=0.5$, $C=2$. The present
dispersions capture the necessary ingredients allowing for the $S_{\pm}$ state to emerge, namely well separated electron and hole Fermi surface (FS) sheets. Moreover, we have
verified that the results that we report here \emph{are independent of any further band structure details}. In fact, we keep our analysis as generic as possible by adopting a
separable potentials approach for the effective interactions, allowing for a broad discussion concerning a large number of iron-based SC.

For this two band system we can write
\begin{widetext}
\bea
{\cal H}=\sum_{\bm{k},\sigma}\left[\varepsilon_{e}(\bm{k}) c_{\bm{k},\sigma}^{\dag}c_{\bm{k},\sigma}\up+\varepsilon_{h}(\bm{k})d_{\bm{k},\sigma}^{\dag}d_{\bm{k},\sigma}\up\right]
-\frac{2}{N}\sum_{\bm{k},\bm{k}'}
V(\bm{k},\bm{k}')\left(c_{\bm{k},\uparrow}^{\dag} c_{-\bm{k},\downarrow}^{\dag}+d_{\bm{k},\uparrow}^{\dag} d_{-\bm{k},\downarrow}^{\dag}\right)
\left(c_{-\bm{k}',\downarrow}\up c_{\bm{k}',\uparrow}\up+d_{-\bm{k}',\downarrow}\up d_{\bm{k}',\uparrow}\up\right)\,,\label{HMK}\eea
\end{widetext}

\noi where $c_{\bm{k},\sigma}^{(\dag)}$ and $d_{\bm{k},\sigma}^{(\dag)}$ are annihilation (creation) operators for the electron $\varepsilon_{e}(\bm{k})$ and hole band
$\varepsilon_{h}(\bm{k})$ bands respectively of spin projection $\sigma=\uparrow,\downarrow$  and $N$ the number of lattice points. Notice that for the specific choice
of the interaction \cite{note}, where intraband and interband interaction strengths are equal, electron and hole bands share the same superconducting order parameter $\Delta(\bm{k})$,
which is defined in the following manner
\bea
\Delta(\bm{k})&=&-\frac{2}{N}\sum_{\bm{k}'}V(\bm{k},\bm{k}')\left<c_{-\bm{k}',\downarrow}\up c_{\bm{k}',\uparrow}\up+d_{-\bm{k}',\downarrow}\up d_{\bm{k}',\uparrow}\up\right>\no\\
\Delta^*(\bm{k})&=&-\frac{2}{N}\sum_{\bm{k}'}V(\bm{k},\bm{k}')\left<c_{\bm{k}',\uparrow}^{\dag} c_{-\bm{k}',\downarrow}^{\dag}+d_{\bm{k}',\uparrow}^{\dag} d_{-\bm{k}',\downarrow}
^{\dag}\right>.\quad
\eea

\noi The order parameter $\Delta(\bm{k})=\sum_n\Delta_nf_n(\bm{k})$ consists of irreducible representations (IR) $f_n(\bm{k})$ of the point group D$_{4h}$, that includes C$_4$ as a
subgroup, and is appropriate for describing the normal (tetragonal) phase of strongly two-dimensional pnictide compounds. Here we shall restrict our analysis to the following IRs:
$f_S(\bm{k})=1$ (A$_{1g}$), $f_{S_{\pm}}(\bm{k})=\cos k_x+\cos k_y$ (A$_{1g}$) and $f_{d_{x^2-y^2}}(\bm{k})=\cos k_x-\cos k_y$ (B$_{1g}$). All of them are even under inversion as
it is required for singlet intraband superconductivity and connect up to nearest neighbors. Note that by Fourier transforming the corresponding effective interaction field
$V(\bm{k},\bm{k}')=\sum_nV_{n}f_{n}(\bm{k})f_{n}(\bm{k}')$ with $n=S,S_{\pm},d_{x^2-y^2}$ one can see that in the context of real space extended Hubbard models this would correspond
to on-site interactions of the form $V_{S}\equiv U\delta_{i,i}$ and nearest neighbor interactions of the form
$V_{S_{\pm}},V_{d_{x^2-y^2}}\equiv V_{i,j}(\delta_{j,i+(\pm 1,0)}+\delta_{j,i+(0,\pm 1)})$, where $i$ and $j$ the real-space lattice points indices. Nevertheless extended Hubbard
models are not the unique option since also a small-q phonon-mediated pairing potential leads to a plethora of non $s$-wave IRs such as $S_{\pm}$ or $d_{x^2-y^2}$
\cite{SmallQprb,AperisSmallQprb2011}. With this reasoning we conclude that by studying here the SC phase competition via separable potentials we report generic results that are
universally relevant, and independent of the exact microscopic mechanism of SC.

Within the aforementioned subspace of IRs, there are only two minimal schemes to achieve a nematic state. These are the mixed states $S+d_{x^2-y^2}$ and $S_{\pm}+d_{x^2-y^2}$. Of
course, the cases $S+S_{\pm}+d_{x^2-y^2}$, $iS+S_{\pm}+d_{x^2-y^2}$, $S+iS_{\pm}+d_{x^2-y^2}$ are also possible but not minimal. In all these symmetry breaking patterns the subgroup
C$_{4}$ reduces to C$_{2}$. Notice that for a minimal nematic phase, the two IRs involved must lock in the same phase. If the two phases lock in phases with $\pi/2$ difference then
the mixed state leads to broken ${\cal T}$ but unbroken C$_4$. These states are also important and for completeness we shall also discuss features of their phase diagram and their
phenomenology.

We consider first the minimal configurations in which nematicity emerges and ${\cal T}$ is preserved. In these cases the order parameter $\Delta(\bm{k})$ is the sum of two order
parameters $\Delta_{1,2}(\bm{k})$ corresponding to the two coexisting IRs. For simplicity we shall consider that $\Delta_{1,2}(\bm{k})$ are real. At this point, we introduce the
spinor $\Psi_{\bm{k}}^{\dag}=(c_{\bm{k},\uparrow}^{\dag} \;d_{\bm{k},\uparrow}^{\dag} \; c_{\bm{-k},\downarrow}\up \; d_{\bm{-k},\downarrow}\up)$ and employ the usual Pauli
matrices $\hat{\tau},\hat{\rho}$. The mean field Hamiltonian can be written in the following compact form
\bea
{\cal{H}} =\sum_{\bm{k}}\Psi_{\bm{k}}^{\dag} \biggl\{\hat{\tau}_{3}\left(\begin{array}{cc}\varepsilon_{e}(\bm{k})& 0 \\0& \varepsilon_{h}(\bm{k})\end{array} \right)-\qquad\qquad
\qquad \nonumber\\
-\mu\hat{\tau}_{3}\hat{\rho}_{0}-{\cal B}\hat{\tau}_{0}\hat{\rho}_{0}+\Delta_{1}(\bm{k})\hat{\tau}_{1}\hat{\rho}_{0}+\Delta_{2}(\bm{k})\hat{\tau}_{1}\hat{\rho}_{0}\biggr\}\Psi_{\bm{k}}
\eea

\noi where we also incorporated the effect of a Zeeman field ${\cal B}$. With usual techniques we calculate Green's functions that exhibit four quasiparticle branches and coupled
self-consistent equations that provide the two gaps $\Delta_{1,2}(\bm{k})$.
\begin{figure}[b]
\includegraphics[scale=0.34]{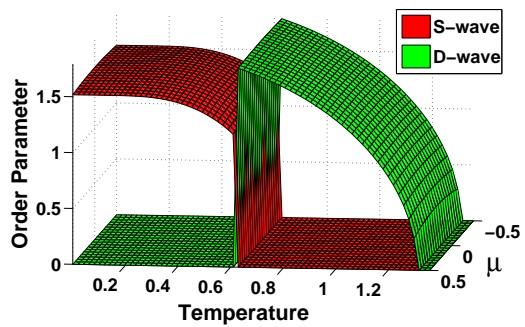}
\caption{(color online) Typical temperature induced first order transition from $S$ (red) to $d_{x^2-y^2}$ SC (green) for various chemical potentials $\mu$ obtained when
$V_{S}<V_{d_{x^2-y^2}}<1.5V_{S}$. Free energy calculations not reported here verify the reality of this transition.}
\end{figure}

To illustrate the fact that achieving mixed states is not trivial, we start with the competition between the isotropic $S$ IR and the $d_{x^2-y^2}$ IR. As expected, s and d-wave SC
phases are highly competitive and {\it coexistence cannot be achieved} at any value of the respective potentials. Remarkably, when $V_{S}<V_{d_{x^2-y^2}}<1.5V_{S}
\label{POT1} $ we observe a {\it first order transition from d-wave to s-wave SC} as we lower the temperature (Fig. 1). We confirmed the reality of this transition, as we did for all
results reported here, by verifying that it minimizes the corresponding \emph{free energy}.

\begin{figure}[t]
\includegraphics[scale=0.32]{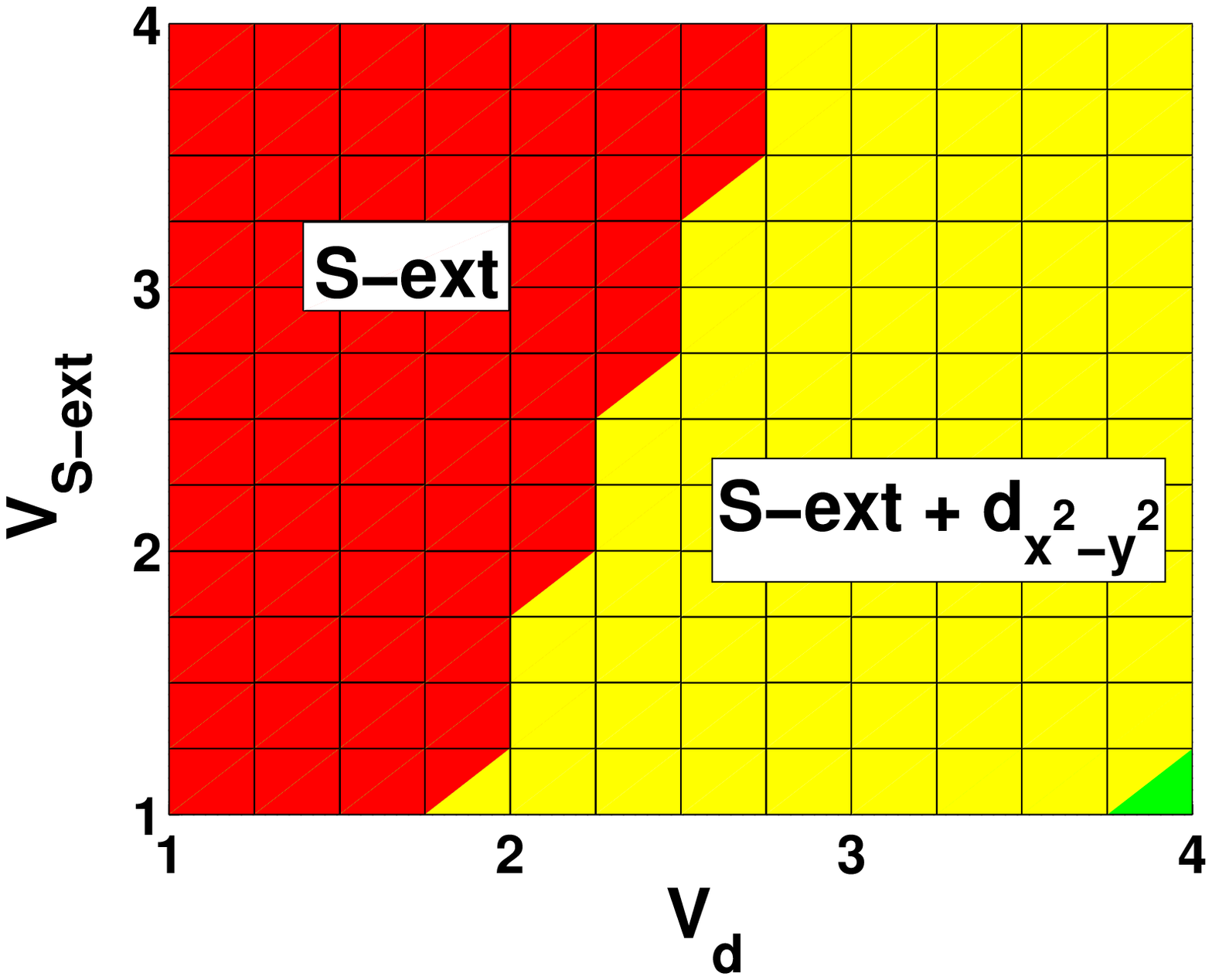}
\includegraphics[scale=0.32]{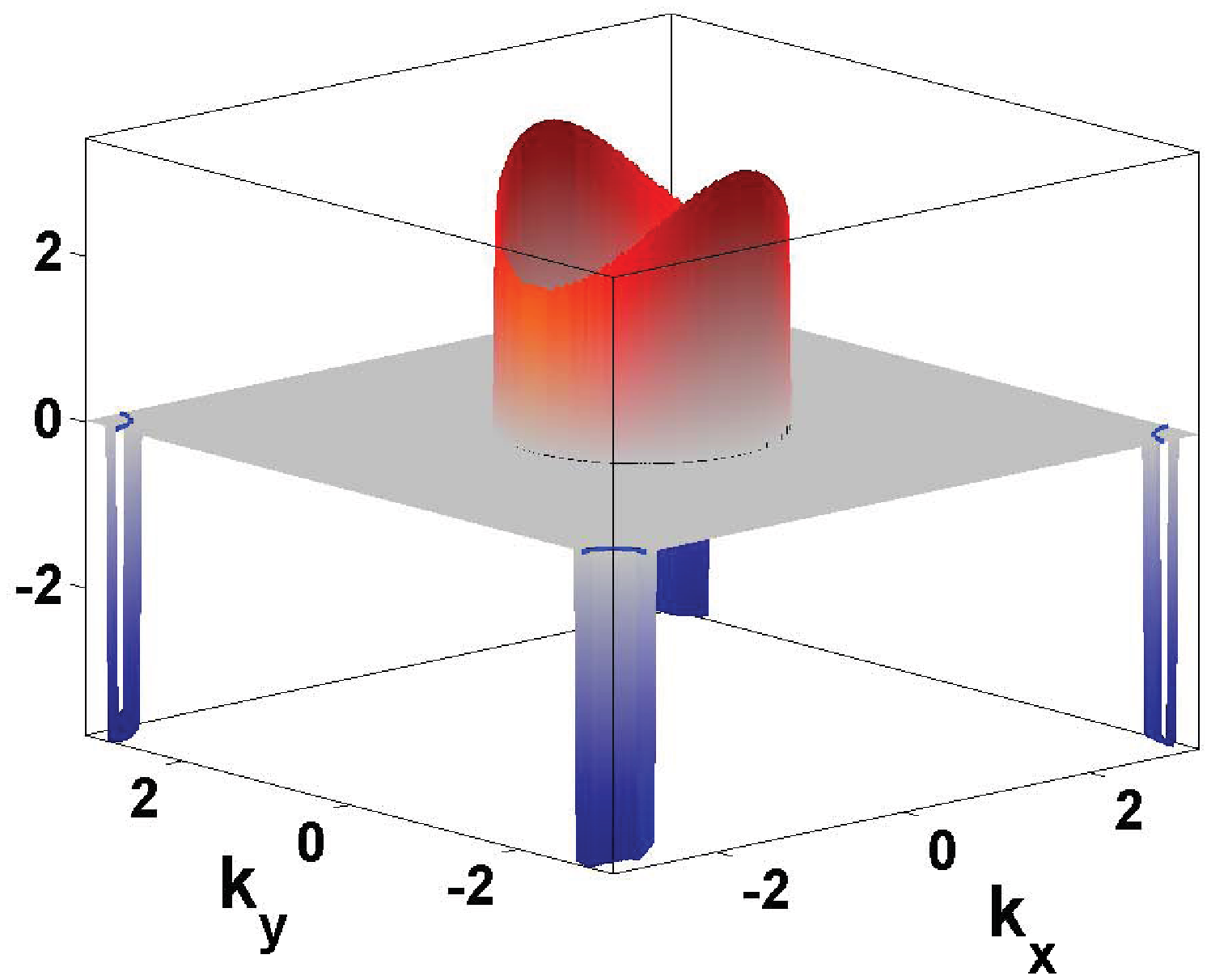}
\includegraphics[scale=0.32]{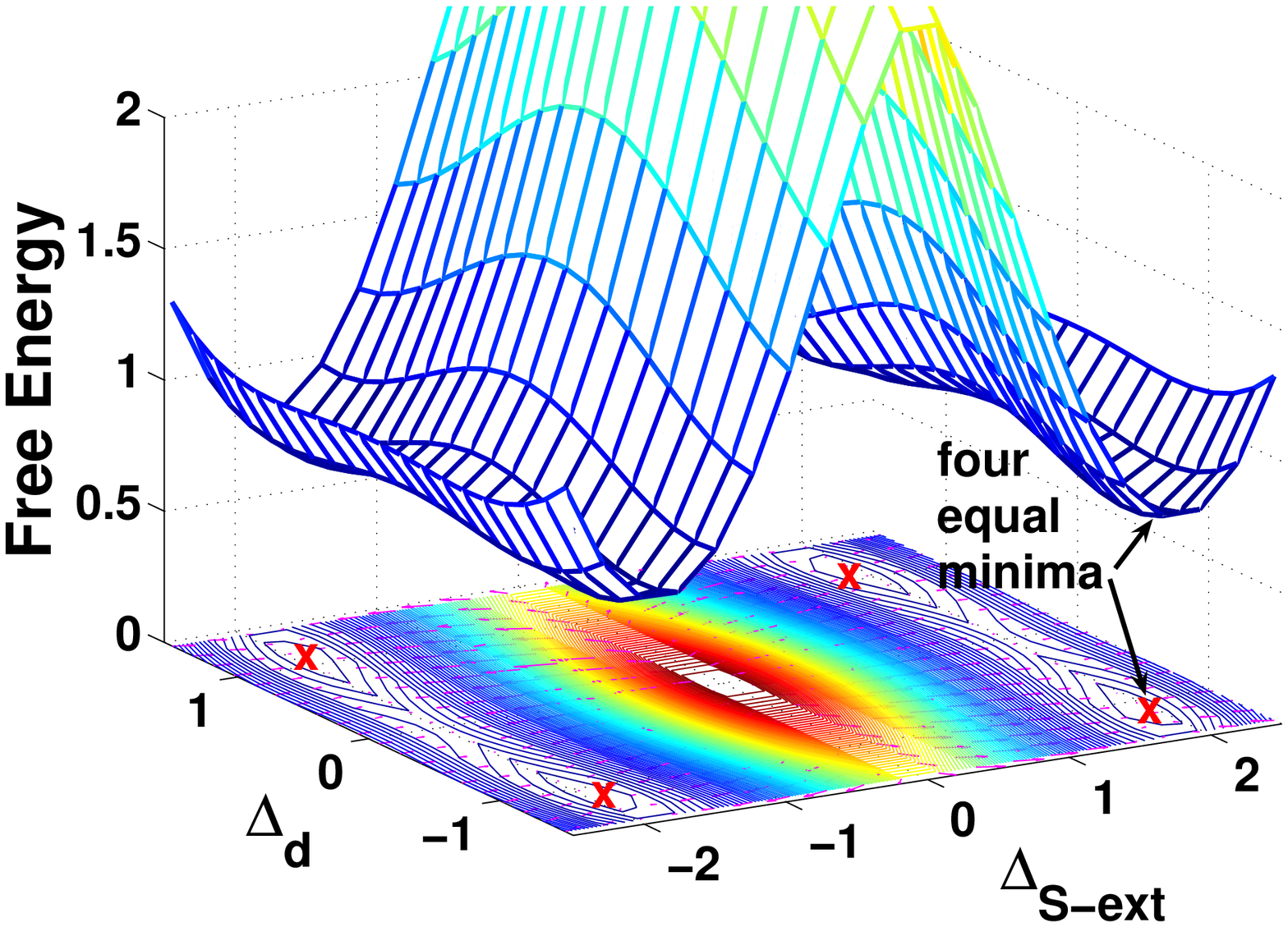}
\caption{(color online) a. Map of the phases obtained varying $V_{S_\pm}\equiv V_{S-ext}$ and $V_{d_{x^2-y^2}}$ for zero temperature in the hole doped ($\mu=-0.4$)
regime ($S_{ext}\equiv S_{\pm}$ (red), $S_{\pm}+d_{x^2-y^2}$ (yellow), $d_{x^2-y^2}$ (green)). Notice the extended region over which $S_{\pm}$ and $d_{x^2-y^2}$ coexist. b. An
example of self-consistently obtained total SC gap amplitude in the nematic $S_{\pm}+d_{x^2-y^2}$ state plotted on the electron and hole Fermi surface sheets for a hole
doped case $\mu=-0.4$. Note that the tetragonal symmetry C$_4$ has been reduced to C$_2$. c. Corresponding free energy results as a function of the $d_{x^2-y^2}$ and $S_{\pm}$ gaps
($\Delta_{d_{x^2-y^2}}$ and $\Delta_{S_{\pm}}$ respectively) exhibiting four degenerate total minima for which both order parameters are finite confirming the $S_{\pm}+d_{x^2-y^2}$
state.}
\end{figure}

While the isotropic s-wave SC state cannot coexist with the d-wave state, the extended s-wave $S_{\pm}$ state that is widely considered relevant for iron-based superconductors
{\it coexists} with $d_{x^{2}-y^{2}}$ over a wide range of the effective potentials (Fig. 2a). We report in Fig. 2b a typical solution in the mixed $S_{\pm}+d_{x^2-y^2}$ state
plotted solely on the Fermi surface for a hole doped system $\mu=-0.4$. The emergence of nematicity is directly evident from Fig. 2b. In
Fig. 2c we present free energy calculations, exhibiting four degenerate (because of symmetry) minima corresponding to the mixed nematic SC state. We insist that all effective
potentials and dispersions used in our self-consistent calculations \emph{preserve tetragonal symmetry}. Only because $S_{\pm}$ and $d_{x^{2}-y^{2}}$ \emph{coexist}, tetragonal
symmetry is broken and nematicity emerges. The essential ingredients leading to the nematic $S_{\pm}+d_{x^2-y^2}$ state is on one hand the well separated electron and hole pockets
that favor the stabilization of the $S_{\pm}$ IR and on the other, some weak tendency towards the formation of the $d$-wave that is further assisted by the presence of $S_{\pm}$.
The detailed characteristics of the Fermi surface topology are not crucial for the formation of the nematic state but mainly determine the exact balance of the $S_{\pm}$ and
$d_{x^2-y^2}$ OPs, that controls the nodal or nodeless type of the quasiparticle excitation spectrum.

\begin{figure}[t]
\includegraphics[scale=0.34]{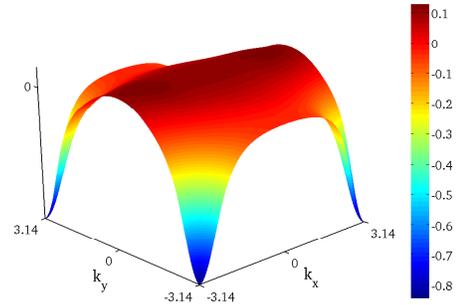}
\caption{(color online)
Self-consistently obtained nodal nematic $S_{\pm}+d_{x^2-y^2}$ state (plotted over the whole Brillouin zone) using a fully momentum dependent small-q electron-phonon pairing potential
and a realistic four band model for hole doped iron pnictides described in Ref. \cite{AperisSmallQprb2011}.}
\end{figure}

To demonstrate that our qualitative findings are not peculiar to details of the considered dispersion or to the separable potentials character of the above analysis, we also report
in Fig. 3 a self-consistently obtained nodal nematic $S_{\pm}+d_{x^{2}-y^{2}}$ state using a fully momentum dependent small-q phonon-mediated interaction $V(\bm{k},\bm{k}')=
V_{Cb}^*-V_{ph}(q_c^2+|\bm{k}-\bm{k}'|^2)^{-1}$ ($V_{Cb}\approx0.1V_{ph}$ and $q_c=\pi/6$) and an accurate hole doped ($\mu=-0.6$) four band model for high-$T_c$ iron
pnictides described in \cite{AperisSmallQprb2011}. Naturally, also in this case both the dispersion and the interaction used in our calculations preserve the tetragonal symmetry, and
only the resulting self-consistent solution depicted in Fig. 3 exhibits nematicity. Note that small-q phonon driven unconventional SC has also been considered in the past
for high-T$_c$ cuprates \cite{SmallQprb,cupratesSmallQ}, heavy fermion \cite{heavyfermionSmallQ} organic \cite{organicSmallQ} and cobaltite \cite{cobaltiteSmallQ} SC and is known to
produce a loss of rigidity of the gap function in momentum space
called {\it momentum decoupling} \cite{SmallQprb} thus
 allowing for gap symmetry transitions. Note also that depending on the relative magnitude of the two order parameters, which in turn depends on the
effective potentials, the mixed nematic state $S_{\pm}+d_{x^{2}-y^{2}}$ can be either nodeless as in the example of Fig. 2b or nodal as in Fig. 3. The small-q results are just a
particular case confirming that the findings of the separable potentials analysis is generic. The nematic $S_{\pm}+d_{x^{2}-y^{2}}$ SC state is indeed a model independent phenomenon
likely to be behind the puzzling reports of nematicity in FeSe films \cite{ScienceFeSe}.

\begin{figure}[t]
\includegraphics[scale=0.38]{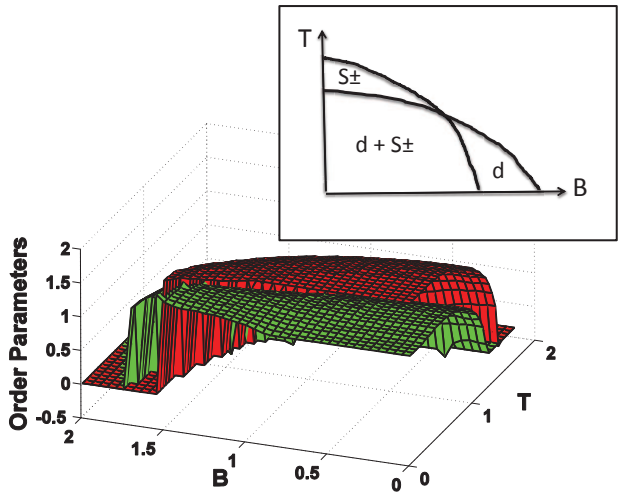}
\caption{(color online) Zeeman field - Temperature phase diagram resulting from our self-consistent calculations for the case shown in Fig. 2b. It exhibits three distinct
superconducting regions and a \emph{tetracritical} point as it is schematically depicted in the inset. A similar diagram has been experimentally observed in UPt$_3$
\cite{SigristUeda}.}
\end{figure}

Moreover, the qualitative behavior of the mixed $S_{\pm}+d_{x^{2}-y^{2}}$ state in the presence of a Zeeman field may allow its firm identification from the experiments. At low
temperatures, we obtain a remarkable \emph{first-order field-induced transition from the nematic $S_{\pm}+d_{x^{2}-y^{2}}$ state to the pure $d_{x^{2}-y^{2}}$
state} (Fig. 4). The transition exists only at sufficiently low temperatures. We therefore propose an experimental approach for identifying the $S_{\pm}+d_{x^{2}-y^{2}}$ state by
applying \emph{in-plane} magnetic fields to the FeSe films instead of perpendicular fields as in \cite{ScienceFeSe}. Exploring the higher temperatures in the presence of the field
allows to construct with \emph{self-consistent calculations} Field - Temperature phase diagrams. Quite remarkably, for the zero field state shown in Fig. 2b, we obtain a
Field-Temperature phase diagram exhibiting \emph{three} distinct SC regions and a \emph{tetracritical} point (Fig. 4) in analogy to the well known phase diagram of UPt$_3$ obtained
there as well in the presence of in-plane fields \cite{SigristUeda}. In our case there is a simple understanding of this diagram as follows: The $S_{\pm}$ state is stronger in this
example owing the higher $T_c$ at zero field. On the other hand, in the presence of a large Zeeman field, the $d_{x^{2}-y^{2}}$ state with nodes on the FS is energetically more favorable
than a nodeless SC state \cite{organicSmallQ}, like $S_{\pm}$, which exhibits a higher critical field at zero temperature. Therefore, the reason for such a complicated phase diagram
lies in the extraordinary fact that at zero field, \emph{a fully gapped state like $S_{\pm}$ allows at lower temperatures its coexistence with an emergent nodal d-wave state}.

\begin{figure}[t]
\includegraphics[scale=0.28]{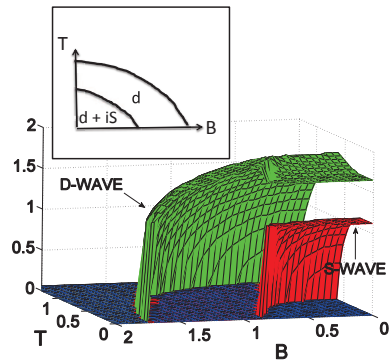}
\caption{(color online) Field and temperature behavior of the mixed $S+id_{x^{2}-y^{2}}$ state obtained for $V_{d_{x^2-y^2}}$ sufficiently stronger than $V_S$. Such a mixed state
breaks time reversal invariance but preserves the tetragonal symmetry. Here as well there is at low-T a first order Zeeman field induced transition from the mixed to the pure
$d_{x^2-y^2}$ state. The inset depicts schematically the corresponding phase diagram.}
\end{figure}

Mixed states may also lead to ${\cal T}$ breaking despite the singlet character of the condensates. This naturally happens if the two coexisting order parameters {\it lock in a
$\pi/2$ phase difference}. As we have already mentioned, when two different IRs coexist with this type of phase locking C$_4$ symmetry is preserved. The mean field Hamiltonian
describing such a situation has the following form
\bea
{\cal{H}} =\sum_{\bm{k}}\Psi_{\bm{k}}^{\dag} \biggl\{\hat{\tau}_{3}\left(\begin{array}{cc}\varepsilon_{e}(\bm{k})& 0 \\0& \varepsilon_{h}(\bm{k})\end{array} \right)-\qquad\qquad
\qquad \nonumber\\
-\mu\hat{\tau}_{3}\hat{\rho}_{0}-{\cal B}\hat{\tau}_{0}\hat{\rho}_{0}+\Delta_{1}(\bm{k})\hat{\tau}_{1}\hat{\rho}_{0}-\Delta_{2}(\bm{k})\hat{\tau}_{2}\hat{\rho}_{0}\biggr\}\Psi_{\bm{k}}
\eea

\noi leading to a different system of coupled self-consistent gap equations, compared to the previously examined case. We briefly report some results in Fig. 5 that will help the
interpretation of eventually observed in-plane field-induced first order transitions in the experiments that we suggest in the previous paragraph. While the state $S+d_{x^2-y^2}$ is
not accessible, breaking of ${\cal T}$ due to $\pi/2$ phase locking, allows for sufficiently large d-wave potentials ($V_{d_{x^2-y^2}} > 1.2V_S$) self-consistent solutions of the
$S+id_{x^2-y^2}$ (or equivalently $iS+d_{x^2-y^2}$) type. Remarkably, here as well, we obtain at zero temperature a first order Zeeman field-induced transition from the mixed
$S+id_{x^2-y^2}$ ($iS+d_{x^2-y^2}$) state to the pure $id_{x^{2}-y^{2}}$ ($d_{x^2-y^2}$) state (Fig 5) that exhibits a qualitatively similar behavior to the field-induced transition
from the $S_{\pm}+d_{x^{2}-y^{2}}$ state to the pure $d_{x^{2}-y^{2}}$ discussed in the previous paragraph. Nevertheless, the mixed $S+id_{x^2-y^2}$ state preserves the tetragonal
symmetry and therefore a \emph{first order Zeeman field-induced transition within the SC phase does not necessarily imply the presence of a nematic SC state}. The Zeeman field -
temperature phase diagram depicted in Fig. 5 exhibits now only two distinct SC regions because the nodal $d_{x^{2}-y^{2}}$ state with the higher critical field is now stronger at
zero field than the $S_{\pm}$ state having the higher critical temperature as well. On the other hand, for the competition of $d_{x^{2}-y^{2}}$ ($id_{x^2-y^2}$) with $iS_{\pm}$
($S_{\pm}$) we observe no qualitative difference in the phase diagrams compared to the one obtained previously when the two order parameters lock in the same phase.
In fact, it has been suggested that enhanced pnictogen height may favor this type of ${\cal T}$ violating mixed states \cite{Platt}.

Finally, from the point of view of BCS theory, unconventional particle-hole condensates like spin singlet or triplet Density Waves (DW) behave similarly to SC condensates when the
two bands are perfectly nested. Qualitatively, our present findings could be extrapolated to the DW condensates suggesting that the nematicity that accompanies the antiferromagnetic
transition in undoped and underdoped iron-pnictides may well indicate the presence of a \emph{nematic mixed $S_{\pm}+d_{x^{2}-y^{2}}$ spin DW state}. Note that in particle-hole
asymmetric systems a spin DW and charge DW exhibiting the same momentum structure coexist \cite{CMR}, and a nematic charge DW may indeed drive the orthorhombic distortion in the form
of a Peierls instability\cite{Analytis}. If it is proven that unconventional SC in iron based SC emerges in the proximity of unconventional density wave phases of the same symmetry,
then a fundamental analogy to high-T$_c$ cuprates emerges where there are reports of a d-wave DW (orbital antiferromagnet) associated with the pseudogap \cite{dDW}. Unconventional
density wave states may host exotic phenomena like emergent chiral density wave states \cite{CDDW} induced by the orbital effects of a magnetic field \cite{FieldChiral} producing
extraordinary phenomena like the anomalous Nernst signal \cite{Kotetes} observed in underdoped cuprates \cite{Ong}. Anomalies in the Nerst signal are apparently present in underdoped
pnictides as well at temperatures where nematicity emerges \cite{NernstPnictides}. Dedicated work is needed in order to substantiate the conjecture of a mixed nematic density waves
state in pnictides.

In conclusion, we claim that nematicity in iron-based superconductors indicates the presence of mixed condensates in which $d_{x^2-y^2}$ and $S_{\pm}$ order parameters coexist. We
suggest the experimental search for an in-plane field-induced melting of nematicity in films of FeSe by an abrupt first order transition at low temperatures. Mixed states that break
time reversal invariance are also shown to be accessible and should be taken into consideration in the analysis of experiments. If our findings for nematic mixed SC states are proven
to be relevant in FeSe films, then it is probable that nematicity in non-superconducting phases of underdoped pnictides originates from analogous mixed nematic density wave
condensates.

We are grateful to Sergey Borisenko, Anna B\"{o}hmer, Christoph Meingast and J\"{o}rg Schmalian for illu\-mi\-nating comments and discussions on the phenomenology of the pnictides.
We acknowledge funding from the $\Pi E B E$ program of NTUA.


\begin{thebibliography}{99}

\bibitem{ScienceFeSe} Can-Li Song {\it et al.}, Science {\bf 332}, 1410 (2011).

\bibitem{nodeless} H. Ding {\it et al.}, Europhys. Lett. {\bf 83}, 47001 (2008); K. Terashima {\it et al.}, Proc. Natl. Acad. Sci. {\bf 106}, 7330 (2009); Y. Zhang {\it et al.},
Nature Mater. {\bf 10}, 273 (2011).

\bibitem{Spm} I. I. Mazin, D. J. Singh, M. D. Johannes and M.H. Du, Phys. Rev. Lett. {\bf 101}, 057003 (2008); V. Stanev, J. Kang and Z. Tesanovic, Phys. Rev. B {\bf 78}, 184509 (2008);
A.V. Chubukov, D.V. Efremov and I. Eremin, Phys. Rev. B {\bf 78}, 134512 (2008).

\bibitem{nodal} J. D. Fletcher {\it et al.}, Phys. Rev. Lett. {\bf 102}, 147001 (2009); J. K. Dong {\it et al.}, Phys. Rev. Lett. {\bf 104}, 087005 (2010); Y. Nakai {\it et al.}, Phys.
Rev. B {\bf 81}, 020503 (2010); M. Yamashita {\it et al.}, arXiv:1103.0885 (2011); Y. Zhang {\it et al.}, arXiv:1109.0229; X. Qiu {\it et al.}, arXiv:1112.2421.

\bibitem{Daghero} D. Daghero, M. Tortello, G. A. Ummarino and R. S. Gonelli, Rep. Prog. Phys. {\bf 74}, 124509 (2011).

\bibitem{Bernevig} R. Thomale, C. Platt, W. Hanke and B.A. Bernevig, Phys. Rev. Lett. {\bf 106}, 187003 (2011); R. Thomale {\it et al.}, Phys. Rev. Lett. {\bf 107}, 117001 (2011).

\bibitem{AperisSmallQprb2011} A. Aperis, P. Kotetes, G. Varelogiannis and P. M. Oppeneer, Phys. Rev. B {\bf 83}, 092505 (2011).

\bibitem{expsTripletLiFeAs} T. H\"{a}nke {\it et al.},Phys. Rev. Lett. {\bf 108}, 127001 (2012).

\bibitem{Fradkin} E. Fradkin, S. A. Kivelson, M. J. Lawler, J.P. Eisenstein and A.P. Mackenzie, Annual Reviews of Condensed Matter Physics {\bf 1}, 153 (2010).

\bibitem{Analytis} J.-H. Chu, H.-H. Kuo, J. G. Analytis and I. R. Fisher, arXiv 1203:3239.

\bibitem{Meingast} C. Meingast \textit{et al}., Phys. Rev. Lett. \bt{108}, 177004 (2012).

\bibitem{Boehmer} A. E. B\"{o}hmer \textit{et al}., arXiv:1203.2119.

\bibitem{Kivelson} C. Fang, H. Yao, W.-F. Tsai, J. Hu and S. A. Kivelson, Phys. Rev. B {\bf 77}, 224509 (2008); J. Hu, C. Setty and S. A. Kivelson, arXiv:1201.5174.

\bibitem{Sachdev} C. Xu, M. Mueller and S. Sachdev, Phys. Rev. B {\bf 78}, 20501 (2008); E.-G. Moon and S. Sachdev, arXiv:1112.3973.

\bibitem{Si} Q. Si and E. Abrahams, Phys. Rev. Lett. {\bf 101}, 076401 (2008); P. Goswami, R. Yu, Q. Si, and E. Abrahams, Phys. Rev. B {\bf 84}, 155108 (2011).

\bibitem{Schmalian} R. M. Fernandez \emph{et al.}, Phys. Rev. Lett. {\bf 105}, 157003 (2010); R. M. Fernandes, A. V. Chubukov, J. Knolle, I. Eremin, and J. Schmalian, Phys. Rev. B
\bt{85}, 024534 (2012); R. M. Fernandes and J. Schmalian, arXiv:1204.3694.

\bibitem{Phillips} W. Lv and P. Phillips, Phys. Rev. B {\bf 84}, 174512 (2011);
H.-H. Hung {\it et al.}, Phys. Rev. B {\bf 85}, 104510 (2012).

\bibitem{FuldeFilms} P. Fulde, Advances in Physics, {\bf 22}, 667 (1973).

\bibitem{SigristUeda} M. Sigrist and K. Ueda, Rev. Mod. Phys. {\bf 63}, 239 (1991); R. Joynt and L. Taillefer, Rev. Mod. Phys. {\bf 74}, 235 (2002).

\bibitem{note} The factor of 2 in front of the potential $V(\bm{k},\bm{k}')$ has been inserted in order to make a connection to self-consistent solutions using four band models of
FeAs compounds \cite{AperisSmallQprb2011}, where the number of electron and hole pockets is doubled. Here we consider that these two extra pockets are identical to the ones
constructed by our minimal two band model, leading to a twofold degeneracy.

\bibitem{Platt} C. Platt {\it et al.}, Phys. Rev. B {\bf 85}, 180502 (2012).

\bibitem{SmallQprb} G. Varelogiannis, Phys. Rev. B {\bf 57}, 13743 (1998).

\bibitem{cupratesSmallQ} A. A. Abrikosov, Phys. Rev. B {\bf 53}, R8910 (1996); {\bf 56}, 446 (1997); G. Varelogiannis, A. Perali, E. Cappelluti, and L. Pietronero, Phys. Rev. B
{\bf 54}, R6877 (1996).

\bibitem{heavyfermionSmallQ} D. F. Agterberg, V. Barzykin, and L. P. Gorkov, Phys. Rev. B {\bf 60}, 14868 (1999); P. M. Oppeneer and G. Varelogiannis, Phys. Rev. B {\bf 68}, 214512
(2003).

\bibitem{organicSmallQ} G. Varelogiannis, Phys. Rev. Lett. {\bf 88}, 117005 (2002).

\bibitem{cobaltiteSmallQ} X.-S. Ye, Z.-J. Yao, and J.-X. Li, J. Phys. Condens. Matter {\bf 20}, 045227 (2008).

\bibitem{CMR} G. Varelogiannis, Phys. Rev. Lett. {\bf 85}, 4172 (2000).

\bibitem{dDW} S. Chakravarty, R. B. Laughlin, D.K. Morr and C. Nayak, Phys. Rev. B {\bf 63}, 094503 (2001).

\bibitem{CDDW} S. Tewari, C Zhang, V. M. Yakovenko and S. Das Sarma, Phys. Rev. Lett. {\bf 100} 217004 (2008); P. Kotetes and G. Varelogiannis, Phys. Rev. B {\bf 78}, 220509(R)
(2008); Europhys. Lett. {\bf 84}, 37012 (2008);  C.-H. Hsu, S. Raghu and S. Chakravarty, Phys. Rev. B {\bf 84}, 155111 (2011).

\bibitem{FieldChiral} J.-X. Zhu and A. V. Balatsky, Phys. Rev. B {\bf 65}, 132502 (2002); P. Kotetes and G. Varelogiannis, Phys. Rev. B {\bf 80}, 212401 (2009).

\bibitem{Kotetes} P. Kotetes and G. Varelogiannis, Phys. Rev. Lett. {\bf 104}, 106404 (2010).

\bibitem{Ong} Yayu Wang, Lu Li, and N. P. Ong, Phys. Rev. B {\bf 73}, 024510 (2006).

\bibitem{NernstPnictides} A. Kondrat, G. Behr, B. B\"{u}chner, C. Hess, Phys. Rev. B {\bf 83}, 092507 (2011).

\end{thebibliography}
\end{document}